\documentclass[twocolumn,times,tighten]{aastex62}

\usepackage{amssymb}
\usepackage{times}
\usepackage{graphicx}
\usepackage{epstopdf}
\usepackage{subfigure}
\usepackage[T1]{fontenc}
\usepackage{aecompl} 
\usepackage{multirow}
\usepackage{rotating}

\usepackage[fleqn]{amsmath}
\usepackage{mathrsfs}
\usepackage{xcolor}

\usepackage{anyfontsize}

\newcommand{\beq}{\begin{equation}}
\newcommand{\eeq}{\end{equation}}

\def\gs{\mathrel{\lower0.6ex\hbox{$\buildrel {\textstyle >}\over{\scriptstyle \sim}$}}}
\def\ls{\mathrel{\lower0.6ex\hbox{$\buildrel {\textstyle <}\over{\scriptstyle \sim}$}}}
\newcommand{\simgt}{\lower.5ex\hbox{$\; \buildrel > \over \sim \;$}}
\newcommand{\simlt}{\lower.5ex\hbox{$\; \buildrel < \over \sim \;$}}

\defcitealias{se+et15_comalit_I}{CoMaLit-I}
\defcitealias{ser+al15_comalit_II}{CoMaLit-II}
\defcitealias{ser15_comalit_III}{CoMaLit-III}
\defcitealias{se+et15_comalit_IV}{CoMaLit-IV}
\defcitealias{xxl_I_pie+al15}{XXL-I}
\defcitealias{xxl_II_pac+al15}{XXL-II}
\defcitealias{xxl_III_gil+al15}{XXL-III}
\defcitealias{xxl_IV_lie+al15}{XXL-IV}
\defcitealias{xxl_XIII_eck+al15}{XXL-XIII}

\defcitealias{ser+al17_CLUMP_M1206}{CLUMP-3D--I}
\defcitealias{ume+al18_CLUMP_II}{CLUMP-3D--II}
\defcitealias{chi+al18_CLUMP_III}{CLUMP-3D--III}

\shorttitle{CLUMP-3D. Testing $\Lambda$CDM}
\shortauthors{Sereno et al.}

\begin{document}

\title{CLUMP-3D. Testing $\Lambda$CDM with galaxy cluster shapes}

\correspondingauthor{Mauro Sereno}
\email{mauro.sereno@oabo.inaf.it}

\author[0000-0003-0302-0325]{Mauro Sereno}
\affil{INAF - Osservatorio di Astrofisica e Scienza dello Spazio di Bologna, via Piero Gobetti 93/3, I-40129 Bologna, Italy\\}
\affil{Dipartimento di Fisica e Astronomia, Universit\`a di Bologna, via Piero Gobetti 93/2, I-40129 Bologna, Italy\\}

\author[0000-0002-7196-4822]{Keiichi Umetsu}
\affiliation{Institute of Astronomy and Astrophysics, Academia Sinica, P. O. Box 23-141, Taipei 10617, Taiwan\\}

\author[0000-0003-4117-8617]{Stefano Ettori}
\affil{INAF - Osservatorio di Astrofisica e Scienza dello Spazio di Bologna, via Piero Gobetti 93/3, I-40129 Bologna, Italy\\}
\affil{INFN, Sezione di Bologna, viale Berti Pichat 6/2, I-40127 Bologna, Italia\\}

\author[0000-0002-8213-3784]{Jack Sayers}
\affil{Division of Physics, Math, and Astronomy, California Institute of Technology, Pasadena, CA 91125\\}

\author{I-Non Chiu}
\affiliation{Institute of Astronomy and Astrophysics, Academia Sinica, P. O. Box 23-141, Taipei 10617, Taiwan\\}

\author[0000-0003-1225-7084]{Massimo Meneghetti}
\affil{INAF - Osservatorio di Astrofisica e Scienza dello Spazio di Bologna, via Piero Gobetti 93/3, I-40129 Bologna, Italy\\}
\affil{INFN, Sezione di Bologna, viale Berti Pichat 6/2, I-40127 Bologna, Italia\\}

\author[0000-0003-2338-5567]{Jes\'us Vega-Ferrero}
\affiliation{IFCA, Instituto de F\'isica de Cantabria (UC-CSIC), Av. de Los Castros s/n, 39005 Santander, Spain}
\affiliation{Department of Physics and Astronomy, University of Pennsylvania, 209 S. 33rd St, Philadelphia, PA 19104, USA}

\author[0000-0002-0350-4488]{Adi Zitrin}
\affiliation{Physics Department, Ben-Gurion University of the Negev, P.O. Box 653, Be'er-Sheva 84105, Israel}



\begin{abstract}
The $\Lambda$CDM model of structure formation makes strong predictions on concentration and shape of DM (dark matter) halos, which are determined by mass accretion processes. Comparison between predicted shapes and observations provides a geometric test of the $\Lambda$CDM model. Accurate and precise measurements needs a full three-dimensional analysis of the cluster mass distribution. We accomplish this with a multi-probe 3D analysis of the X-ray regular CLASH (Cluster Lensing And Supernova survey with Hubble) clusters combining strong and weak lensing, X-ray photometry and spectroscopy, and the Sunyaev-Zel'dovich effect. The cluster shapes and concentrations are consistent with $\Lambda$CDM predictions. The CLASH clusters are randomly oriented, as expected given the sample selection criteria. Shapes agree with numerical results for DM-only halos, which hints at baryonic physics being not so effective in making halos rounder.
\end{abstract}

\keywords{galaxies: clusters: general --- gravitational lensing: weak --- galaxies: clusters: intracluster medium}


\section{Introduction}
\label{sec_intro}

Cold Dark Matter (CDM) and the cosmological constant $\Lambda$ are well established in modern astrophysics \citep{pee15}. Their case rests on precise measurements on very large scales, notably the cosmic microwave background radiation \citep{planck_2015_XIII} and the baryon acoustic oscillations in the power spectrum of the matter distribution \citep{xu+al13}.


The $\Lambda$CDM paradigm has been very successful and occasional crises have been solved.  At small scales, the properties of the innermost regions of DM halos (the cusp-core problem) or the Milky Way's dwarf galaxy satellites (the missing satellites problem) can be reconciled with $\Lambda$CDM by a better understanding of baryonic physics or deeper observations \citep{wei+al15}. The over-concentration problem and the very large Einstein rings in galaxy clusters can be due to selection effects and statistical biases \citep{og+bl09,men+al11,ser+al15_cM}.

Here, we propose a new test of $\Lambda$CDM based on the shape of cluster-sized halos. A three-dimensional analysis of galaxy clusters beyond the usual spherical hypothesis is crucial in modern astrophysics \citep{lim+al13}.  Halo shapes are the products of structure formation and evolution over billions of years. The matter aggregation from large-scale perturbations determines the shape. The major axes of galaxies and clusters often share the same orientation as the surrounding matter distribution \citep{wes94,ji+su02} possibly since very early epochs \citep{wes+al17}.

Furthermore, we have to account for halo shape and orientation to get unbiased estimates of mass and concentration \citep{ogu+al05,mor+al12,ser+al13}, which are needed in precision cosmology exploiting formation and growth of galaxy clusters.

Shape measurements  are very challenging and demand precise and accurate analyses. Shapes of individual clusters can be recovered with multi-probe approaches \citep{fo+pe02,ma+ch11,mor+al12,tch+al16}. The CLUMP-3D (CLUster Multi-Probes in Three Dimensions) project exploits rich data-sets to infer unbiased measurements of mass and concentration together with the intrinsic shape and the equilibrium status of the cluster \citep{ser+al17_CLUMP_M1206,ume+al18_CLUMP_II,chi+al18_CLUMP_III}. The full three-dimensional Bayesian analysis combines strong (SL) and weak lensing (WL), X-ray surface brightness and temperature, and the Sunyaev-Zel'dovich effect (SZe) \citep{ser+al17_CLUMP_M1206}. In a nutshell, lensing constrains the 2D mass and concentration which are deprojected thanks to the shape and orientation information from X-ray and SZe.

Here, we apply the method to the CLASH (Cluster Lensing and Supernova Survey with Hubble) clusters \citep{pos+al12} and test if the recovered concentrations, shapes and orientations are in agreement with the $\Lambda$CDM predictions. This offers a novel, geometric check of the structure formation and evolution.
 

The reference cosmological model we test is the concordance flat $\Lambda$CDM universe with matter density parameter $\Omega_\text{M}=1-\Omega_\Lambda=0.3$, Hubble constant $H_0=100h~\mathrm{km~s}^{-1}\mathrm{Mpc}^{-1}$ with $h=0.7$, and power spectrum amplitude $\sigma_8=0.82$. 
Notations and conventions follows \citet{ser+al17_CLUMP_M1206}.






\section{The CLASH sample}
\label{sec_samp}

\begin{table*}
\caption{The cluster sample. Columns~4 and 5: right ascension and declination in degrees (J2000) of the associated brightest cluster galaxy (BCG),  adopted as the cluster center. The X-ray data set is detailed in cols.~6--8. Col.~6: observation identification. Col.~7: nominal exposure time. Col.~8: Galactic absorption.}
\label{tab_sample}
\centering
\begin{tabular}{ l l l r r l r c}     
\hline
\multicolumn{2}{c}{Name}		&	$z$  & 	RA	& DEC & ObsID & $t_\text{exp}$ & $n_\text{H}$ \\
Full			&	Short	& &  & 	 & & $[\text{ks}]$ & $[10^{20}\text{cm}^{-2}]$ \\
	\hline
ABELL 0383       	&	A383  	&	0.188	&	42.01409 	&	$-$3.5292641 	& 522/3579 (ACIS-I) & 19.6 & 1.6 \\
ABELL 0209       	&	A209  	&	0.206	&	22.96895 	&	$-$13.6112720	& 524/2320 (ACIS-I) & 28.2 & 3.5 \\
ABELL 2261       	&	A2261 	&	0.225	&	260.61336	&	32.1324650   	&  3194 (ACIS-S) & 32.0 & 4.5 \\
RX J2129+0005    	&	R2129 	&	0.234	&	322.41649	&	0.0892232    	& 5007 (ACIS-I) & 24.1 & 3.2 \\
ABELL 0611       	&	A611  	&	0.288	&	120.23674	&	36.0565650   	& 3257/3582/6108/7719 (ACIS-I) & 71.3 & 4.6 \\
MS 2137.3-2353   	&	MS2137	&	0.313	&	325.06313	&	$-$23.6611360	& 3271 (ACIS-I) & 22.6 & 4.3 \\
RXC J2248.7-4431 	&	R2248 	&	0.348	&	342.18322	&	$-$44.5309080	& 3585/6111 (ACIS-I) & 66.2 & 5.7 \\
MACS J1115+0129  	&	M1115 	&	0.352	&	168.96627	&	1.4986116    	& 3275/9375 (ACIS-I) & 49.1 & 4.3 \\
MACS J1931.8-2635	&	M1931 	&	0.352	&	292.95608	&	$-$26.5758570	& 3277 (ACIS-I) & 22.9 & 3.7 \\
RX J1532.8+3021  	&	R1532 	&	0.363	&	233.22410	&	30.3498440   	& 3280/6107 (ACIS-I) & 51.9 & 3.4 \\
MACS J1720.3+3536	&	M1720 	&	0.391	&	260.06980	&	35.6072660   	& 3282/9382 (ACIS-I) & 111.2 & 8.3 \\
MACS J0429.6-0253	&	M0429 	&	0.399	&	67.40003 	&	$-$2.8852066 	& 5250 (ACIS-S) & 18.9 & 3.8 \\
MACS J1206.2-0847	&	M1206 	&	0.44 	&	181.55065	&	$-$8.8009395 	& 3592/13516/13999 (ACIS-I) & 149.8 & 4.6 \\
MACS J0329.6-0211	&	M0329 	&	0.45 	&	52.42320 	&	$-$2.1962279 	& 14009 (ACIS-S) & 85.8 & 2.3 \\
RX J1347.5-1145  	&	R1347 	&	0.451	&	206.87756	&	$-$11.7526100	& 552/9370 (ACIS-I) & 39.6 & 3.6 \\
MACS J0744.9+3927	&	M0744 	&	0.686	&	116.22000	&	39.4574080   	& 4966/18611/18818 (ACIS-I) & 123.1 & 1.2 \\
\hline	
\end{tabular}
\end{table*}

CLASH is a Multi-Cycle Treasury program with the {\it Hubble Space Telescope} (HST) complemented with high-quality, multi-wavelength data-sets  \citep{pos+al12}. Twenty massive clusters are X-ray selected over the redshift range $0.2\la z \la 0.9$ on the basis of their high temperature $(kT > 5~\text{keV})$, and symmetric and smooth X-ray emission. Five additional clusters are included for their lensing strength to find magnified high-$z$ galaxies. 

We extend the method first applied to M1206 in \citet{ser+al17_CLUMP_M1206} to the 16 X-ray regular CLASH clusters with high quality, ground-based data for WL (Table~\ref{tab_sample}). We do not consider the five lensing selected clusters, which are mostly merging or irregular systems. In fact, our modelling requires that matter and gas follow an ellipsoidal geometry and that gas and matter are aligned and co-centered. Modeling of complex distributions could require using finite mixture models as collections of ellipsoids that fit individual subclusters \citep{kuh+al14}.

The data-sets have been comprehensively presented and detailed elsewhere. In the following, we provide the main references, and detail any change with respect to \citet{ser+al17_CLUMP_M1206}.

{\it  Weak lensing at large radii}. The lensing analysis relies on ground-based data from the Suprime-Cam at Subaru Telescope or the ESO Wide Field Imager \citep{ume+al14,ume+al16,mer+al15}. We mainly refer to \citet{ume+al18_CLUMP_II}, where projected mass maps are recovered from the joint analysis of shear and magnification bias. Our fitting analysis follows \citet{ser+al17_CLUMP_M1206} but we do not limit the fit to a squared region of size of $4~\text{Mpc}/h$, instead we consider the full field of view \citep{chi+al18_CLUMP_III}. Accordingly, the noise from the large-scale structure is added to the uncertainty covariance matrix.

{\it Strong and weak lensing in the cluster cores}. Multiple image systems, shear in the {\it HST} field, and mass models of the inner cluster regions are presented in \citet{zit+al15}. As in \citet{ser+al17_CLUMP_M1206}, we compute the mean convergence from the `PIEMDeNFW' maps in equally spaced circular annuli. The innermost and the outermost radii are set to an angular scale of 5\arcsec and two times the Einstein radius, $ \theta_\text{max, SL}=2 \theta_\text{E}(z_\text{s} = 2)$, respectively. The width of the annuli, $\Delta \theta_\text{SL}$, is determined through the relation $N_\text{im}  \Delta \theta_\text{SL}^2 \sim \pi \theta_\text{E}^2$, where $N_\text{im}$ is the number of images \citep{ume+al16}. We fix $N_\text{im}= N_\text{SL}/2$, where $N_\text{SL}$ is the number of effective SL constraints \citep[table 1]{zit+al15}. The bin size is rounded to have equally spaced annuli. 

{\it X-ray.} Gas density and temperature profiles are measured from archival {\it Chandra} data, see Table~\ref{tab_sample}. The data are analysed as in \citet{ser+al17_CLUMP_M1206}. Cleaned (by grade, status, bad pixels, and time intervals affected from flares in the background count rate) events file are prepared with the CIAO 4.8 software\footnote{\url{http://cxc.harvard.edu/ciao/}} and the calibration database CALDB 4.7.1. Backgrounds are extracted far from the cluster X-ray peak in circular regions of radius of $2\arcmin$. 

Exposure-corrected images are produced in the [0.7-2] keV band. The point-sources are identified with the tool {\tt wavedetct}, filtered out, and the corresponding regions filled with values of counts from surrounding background areas through the tool {\tt dmfilth}. 

We perform the 2D analysis of the number counts in the circular region enclosing  80\% of the total source emission. Pixels are binned with a final resolution of $1.968\arcsec$. We excise the inner region of radius $5\arcsec$. Outside the 80\% region, we examine the surface brightness profiles in circular annuli.

Spectra are extracted in circular annuli and analysed with the XSPEC v.12.9 software\footnote{\url{https://heasarc.gsfc.nasa.gov/xanadu/xspec/}} with an absorbed thermal model {\tt tbabs}, the Galactic absorption fixed by extrapolation from HI radio maps in \citet{kal+al05} and the {\tt apec} emission spectrum with three free parameters (normalisation, temperature and metallicity). The same model with metallicity fixed to the median value is used for regression.

{\it SZe.} The CLASH clusters are part of the Bolocam X-ray SZ (BoXSZ) sample, with publicly available data from Bolocam. Details of the observational campaign and data reduction can be found in \citet{say+al11,say+al16,cza+al15}. Additional data from the {\it Planck} all-sky survey are employed to set the mean signal estimates \citep{ser+al17_CLUMP_M1206}. 

Our analysis strictly follows \citet{ser+al17_CLUMP_M1206}. The integrated Compton parameter is computed from the unfiltered maps in five equally spaced annular bins up to a maximum radius of $5\arcmin$. The annulus width is set to $1\arcmin$, comparable to the PSF FWHM.

\section{The model}
\label{sec_mode}

The parametric joint analysis of the multi-probe data-sets is introduced in \citet{ser+al17_CLUMP_M1206}. The main assumption is that the total matter distribution is ellipsoidal. The halo shape is described by the two axis ratios: $q_{\text{mat},1}$ is the minor-to-major axis ratio, and $q_{\text{mat},2}$  is the intermediate-to-major axis ratio. The orientation is fixed by three Euler's angles: $\vartheta$ is the inclination angle between the major axis and the line-of-sight.

The gas distribution is assumed to be ellipsoidal too and co-centered and co-aligned with the total matter. The gas is taken to be more spherical than the total matter, as usual in regular systems, but we do not require that the gas is in equilibrium in the potential well. This modelling is supported by the analysis of the 2D maps \citep{ume+al18_CLUMP_II}, which show that gas and total matter have a negligible off-set and are aligned in projection, and that a constant matter ellipticity as a function of the radius provides a good description of the data.

WL and X-ray/SZ data probe dark matter and gas, respectively, on different scales. Whereas the matter shape is measured within the viral region, the gas shape is mostly sensitive to the inner regions \citep{ser+al17_CLUMP_M1206}.


We fit the SL and WL convergence maps, the X-ray surface brightness and temperature, and the integrated Compton parameter. The total matter distribution, the gas density, and the gas temperature are modelled with flexible parametric ellipsoidal 3D profiles. Since we do not require hydrostatic equilibrium, the determinations of the matter or the gas density profiles are largely independent apart from an overall normalization related to the orientation \citep{ser+al17_CLUMP_M1206}. The more elongated the cluster is along the line-of-sight, as mainly inferred from X-ray and SZe measurements, the smaller the central gas density, and the smaller the total mass and concentration. Here, we are interested in the global shape and we do not discuss the gas properties, which are simultaneously fitted in the same regression procedure. To conservatively deal with parameter degeneracy, we focus on 1D probability distributions obtained after marginalization of the remaining parameters.


The total mass (DM plus galaxies plus gas) is described as a Navarro-Frenk-White (NFW) density profile  \citep{nfw96},
\begin{equation}
\label{eq_nfw_1}
	\rho_\text{NFW}=\frac{\rho_\text{s}}{(\zeta/\zeta_\text{s})(1+\zeta/\zeta_\text{s})^2},
\end{equation}
where $\zeta$ is the ellipsoidal radius and $\zeta_\text{s}$ is the scale radius. In our notation, $M_\Delta$ is the mass within the ellipsoid of semi-major axis $\zeta_\Delta$, 
\beq
\label{eq_nfw_2}
M_\Delta\equiv(4\pi/3)\Delta\ \rho_\text{cr}(z)\ q_{\text{mat},1}q_{\text{mat},2} \zeta_\Delta^3,
\eeq 
where $ \rho_\text{cr}(z)$ is the critical density. The concentration is $c_\Delta \equiv \zeta_\Delta/ \zeta_\text{s}$. For comparison with numerical simulations, we also consider the mass $M_{\text{sph},\Delta}$ measured in spherical regions, and the concentration $c_{\text{sph},\Delta}$ computed by fitting the spherically averaged NFW profile.

The relation between $\rho_\text{s}$ and the concentration takes the same form in the spherical or ellipsoidal model, whereas $r_\text{sph,s} \sim (q_{\text{mat},1}q_{\text{mat},2})^{1/3} \zeta_\text{s}$. Then, $c_{\text{sph},\Delta}<c_\Delta$.

For our Bayesian analysis, we adopt priors spanning large parameter regions. For mass and concentration, priors are uniform distributions in logarithmically spaced intervals, as suitable for positive parameters \citep{se+co13}: $p_\text{prior}(M_{200})\propto 1/M_{200}$ and $p_\text{prior}(c_{200})\propto 1/c_{200}$ in the allowed ranges $0.01 \le M_{200}/(10^{15}h^{-1}M_\odot) \le 10$ and $1 \le c_{200} \le 10$, or null otherwise. We assume a flat prior for the matter shape ($q$-flat), i.e. the probability $p_\text{prior}(q_{\text{mat},1})$ and the conditional probability $p_\text{prior}(q_{\text{mat},2}|q_{\text{mat},1})$ are constant. The minimum axis ratio is $q_\text{min}=0.1$, and $q_\text{min}\le q_{\text{mat},1} \le q_{\text{mat},2} \le 1$. This prior allows either very triaxial clusters ($q_{\text{mat},1}, q_{\text{mat},2} \ll1$) or spherical systems ($q_{\text{mat},1} \la 1$), which are preferentially excluded by $N$-body simulations \citep{ji+su02}. A priori, the cluster orientation is random, i.e. $p_\text{prior}(\cos \vartheta) = 1$.


\section{Theoretical predictions}

Numerical simulations can picture the cluster properties in the $\Lambda$CDM scenario. We consider two kinds of simulated samples: $i$) halos selected as actual CLASH clusters (`MUSIC2-CLASH'); $ii)$ a complete population of relaxed massive halos (`$\Lambda$CDM-rel').

{\it MUSIC2-CLASH}. Simulated clusters mimicking the CLASH sample are presented in \citet{men+al14}, which study nearly 1400 halos over $0.25\le z\le 0.67$ from the MUSIC-2 $N$-body/hydrodynamical simulations\footnote{\url{http://music.ft.uam.es/}}. These halos are mass-limited ($>10^{15}M_\odot/h$ at $z = 0$) and are re-simulated by adding baryons to the DM distributions \citep{sem+al13}. Here, we consider the runs not including radiative processes.

\citet{men+al14} classified halos as regular or relaxed, the two conditions being non-equivalent. Regular clusters are so in their X-ray features. They show small centroid shift, small ellipticity, and small third- and fourth-order power ratios of the X-ray morphology in the soft-energy band, but large X-ray surface-brightness concentrations. On the other hand, clusters are classified as relaxed according to their center of mass displacement, their virial ratio, and their substructure mass fraction \citep{net+al07}. Regular clusters can be unrelaxed. The X-ray morphology is mostly evaluated in the inner regions, whereas relaxation is evaluated on scales up to the virial radius. 

\citet{men+al14} find simulated avatars of the CLASH X-ray-selected clusters by matching the X-ray morphology. The association does not use gas temperatures or X-ray luminosities, whose physical processes are more challenging to simulate. Masses and redshifts have to be compatible too.

Only the simulated halos closest to each individual CLASH cluster in the morphological parameter space are used. The total number of associations analysed here is 166, with from 2 to 26 associations per cluster. CLASH clusters are found to be prevalently relaxed and only modestly affected by the strong lensing bias. The regularity of the matched clusters is not extreme \citep{men+al14}. The fraction of $\sim 70$ per cent of relaxed halos among X-ray regular clusters is larger than in the full sample, and the average concentration is boosted.

We measure the shapes and orientations of the total matter distribution, i.e. DM plus gas particles, by mimicking our real measurement process \citep{bon+al15}. We compute the mass tensor of the particles selected inside the ellipsoid, centred in the most bound particle, that encloses an overdensity $\Delta =200$. The procedure was reiterated until both $q_{\text{mat},1}$  and $q_{\text{mat},2}$ converge within a 0.5\% of error.


{\it $\Lambda$CDM-rel}. Relaxed galaxy clusters in $N$-body simulations are well represented as a population of ellipsoidal, co-aligned, triaxial halos \citep{ji+su02,all+al06,bon+al15,veg+al17}. Results from different groups agree if methodological differences on how the halo shape is measured are taken into account \citep{veg+al17}. 

We base the $\Lambda$CDM prediction for halo shape on \citet{bon+al15}, who analyse the relaxed halos from the Millennium XXL simulation and provide statistically significant predictions for masses above $3\times10^{14}M_\odot/h$. They measure the shape of the ellipsoid enclosing an overdensity equal to the virial one. Unrelaxed clusters are removed by selecting only halos whose offset between the most bound particle and the centre of mass of the particles enclosed by the ellipsoid is less than 5 per cent of their virial radius. This criterion is not very stringent and a fraction of unrelaxed clusters might still be included.



Theoretical estimates provide a consistent picture of the halo concentrations \citep{bha+al13,du+ma14,lud+al16}. As reference prediction, we follow \citet{men+al14}, who measure the mass-concentration relation under different selection criteria and for either projected or 3D concentrations and masses. The MUSIC-2 halos follow an intrinsic concentration-mass relation with a slightly larger normalisation than other recent results, but with the usual weak redshift evolution. The more sensible comparison to our analysis is with the NFW fitting in 3D of the relaxed sample.

We assume that the unbiased population of relaxed clusters is randomly oriented.

\section{Results}
\label{sec_resu}

\begin{table*}
\caption{Halo properties. $M_{\text{sph,}200}$ (col. 2) and $c_{\text{sph,}200}$ (col.~3) refer to mass and concentration measured in spheres. $M_{200}$ (col. 4) and $c_{200}$ (col.~5) refer to ellipsoids. Masses are in units of $10^{15}~M_\odot/h$. We quote the bi-weighted estimators of the marginalised posterior distributions.}
\label{tab_results}
\centering
\begin{tabular}{ l r@{$\,\pm\,$}l r@{$\,\pm\,$}l r@{$\,\pm\,$}l r@{$\,\pm\,$}l r@{$\,\pm\,$}l r@{$\,\pm\,$}l r@{$\,\pm\,$}l}     
\hline
Cluster			&	\multicolumn{2}{c}{$M_{\text{sph,}200}$}	& \multicolumn{2}{c}{$c_{\text{sph,}200}$}&	\multicolumn{2}{c}{$M_{200}$}	& \multicolumn{2}{c}{$c_{200}$}&	\multicolumn{2}{c}{$q_{\text{mat},1}$}	&	\multicolumn{2}{c}{$q_\text{mat,2}$}	&	\multicolumn{2}{c}{$\cos \vartheta$}	\\
	\hline
A383  	&	0.56   	&	0.11       	&	5.87   	&	0.96       	&	0.63	&	0.14    	&	6.75	&	1.20    	&	0.30 	&	0.06     	&	0.63 	&	0.14     	&	0.25    	&	0.15        	\\
A209  	&	0.80   	&	0.28       	&	1.66   	&	0.45       	&	1.02	&	0.25    	&	1.83	&	0.42    	&	0.28 	&	0.12     	&	0.51 	&	0.24     	&	0.97    	&	0.03        	\\
A2261 	&	1.63   	&	0.28       	&	5.13   	&	0.61       	&	2.06	&	0.37    	&	6.94	&	0.92    	&	0.17 	&	0.04     	&	0.71 	&	0.17     	&	0.40    	&	0.19        	\\
R2129 	&	0.49   	&	0.13       	&	4.55   	&	0.92       	&	0.55	&	0.16    	&	5.17	&	1.09    	&	0.32 	&	0.08     	&	0.68 	&	0.13     	&	0.39    	&	0.17        	\\
A611  	&	0.62   	&	0.15       	&	2.91   	&	0.74       	&	0.75	&	0.18    	&	3.47	&	0.88    	&	0.28 	&	0.12     	&	0.74 	&	0.21     	&	0.94    	&	0.03        	\\
MS2137	&	0.66   	&	0.16       	&	3.60   	&	0.92       	&	0.71	&	0.17    	&	3.78	&	1.07    	&	0.51 	&	0.14     	&	0.85 	&	0.12     	&	0.53    	&	0.19        	\\
R2248 	&	1.16   	&	0.26       	&	4.38   	&	1.06       	&	1.30	&	0.29    	&	4.93	&	1.33    	&	0.34 	&	0.08     	&	0.58 	&	0.16     	&	0.18    	&	0.14        	\\
M1115 	&	1.32   	&	0.27       	&	3.77   	&	0.67       	&	1.63	&	0.34    	&	4.64	&	0.90    	&	0.22 	&	0.04     	&	0.58 	&	0.13     	&	0.20    	&	0.13        	\\
M1931 	&	0.53   	&	0.14       	&	5.51   	&	1.71       	&	0.57	&	0.16    	&	5.94	&	1.68    	&	0.47 	&	0.13     	&	0.72 	&	0.19     	&	0.71    	&	0.26        	\\
R1532 	&	0.51   	&	0.13       	&	6.00   	&	1.08       	&	0.58	&	0.16    	&	7.12	&	1.39    	&	0.30 	&	0.09     	&	0.77 	&	0.16     	&	0.25    	&	0.21        	\\
M1720 	&	0.73   	&	0.17       	&	5.84   	&	1.45       	&	0.80	&	0.18    	&	6.62	&	1.74    	&	0.36 	&	0.09     	&	0.69 	&	0.14     	&	0.41    	&	0.21        	\\
M0429 	&	0.51   	&	0.10       	&	7.03   	&	1.56       	&	0.54	&	0.11    	&	7.58	&	1.69    	&	0.45 	&	0.09     	&	0.85 	&	0.13     	&	0.39    	&	0.25        	\\
M1206 	&	1.20   	&	0.15       	&	6.14   	&	0.60       	&	1.33	&	0.19    	&	6.86	&	0.84    	&	0.33 	&	0.07     	&	0.69 	&	0.13     	&	0.20    	&	0.14        	\\
M0329 	&	0.83   	&	0.16       	&	4.34   	&	0.89       	&	0.93	&	0.22    	&	4.92	&	0.97    	&	0.33 	&	0.10     	&	0.58 	&	0.19     	&	0.65    	&	0.22        	\\
R1347 	&	2.45   	&	0.46       	&	4.61   	&	0.82       	&	3.23	&	0.61    	&	6.34	&	1.33    	&	0.15 	&	0.03     	&	0.66 	&	0.24     	&	0.50    	&	0.19        	\\
M0744 	&	1.23   	&	0.38       	&	4.36   	&	1.09       	&	1.76	&	0.53    	&	6.55	&	1.69    	&	0.11 	&	0.01     	&	0.46 	&	0.17     	&	0.16    	&	0.12        	\\
\hline	
\end{tabular}
\end{table*}

\begin{figure*}
\begin{center}
$
\begin{tabular}{c c}
\noalign{\smallskip}
 \resizebox{0.48\hsize}{!}{\includegraphics{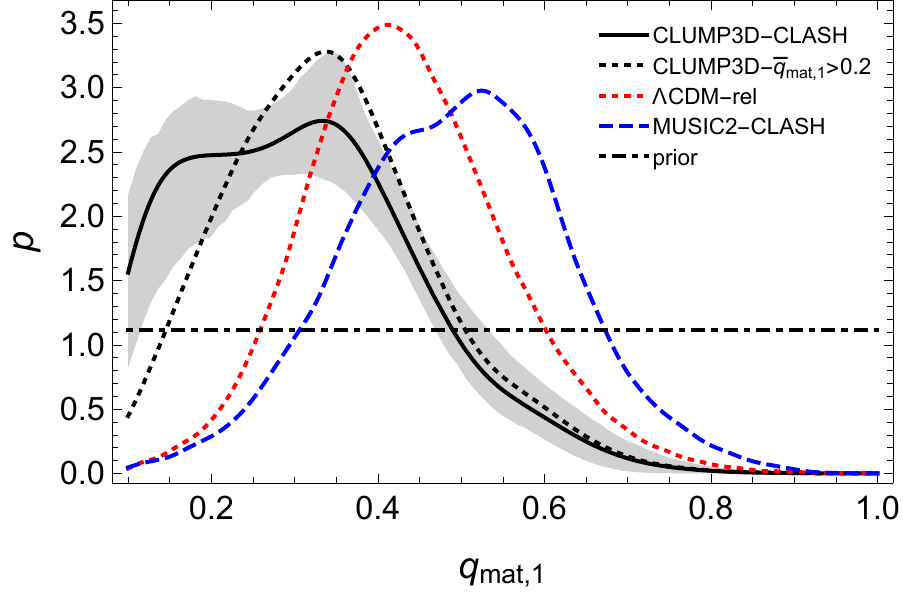}} &  
 \resizebox{0.48\hsize}{!}{\includegraphics{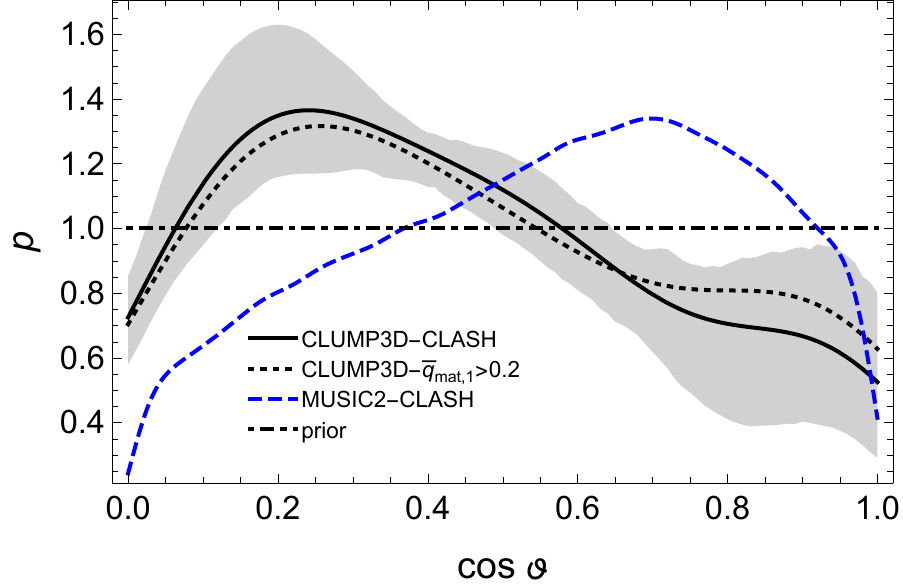}} \\
 \resizebox{0.48\hsize}{!}{\includegraphics{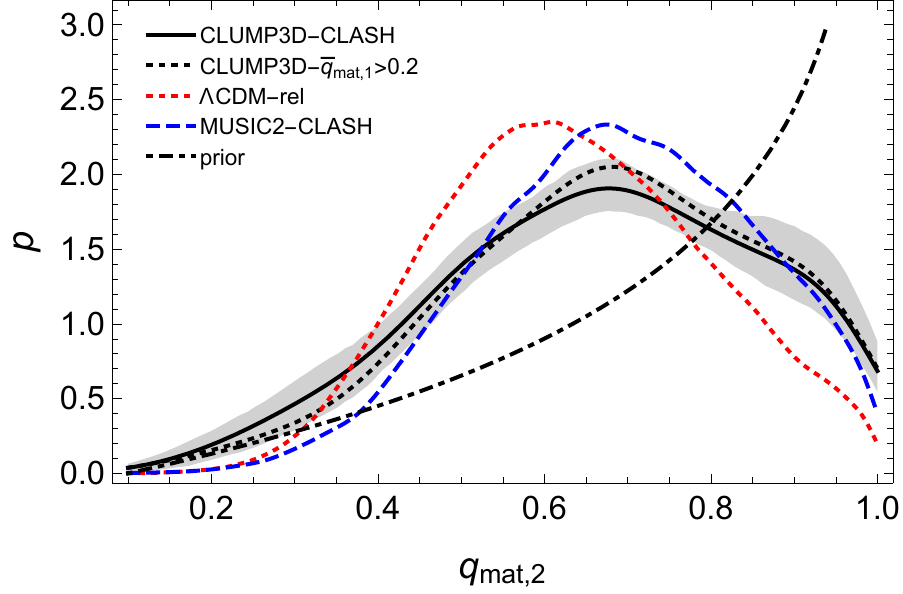}} &  
 \resizebox{0.48\hsize}{!}{\includegraphics{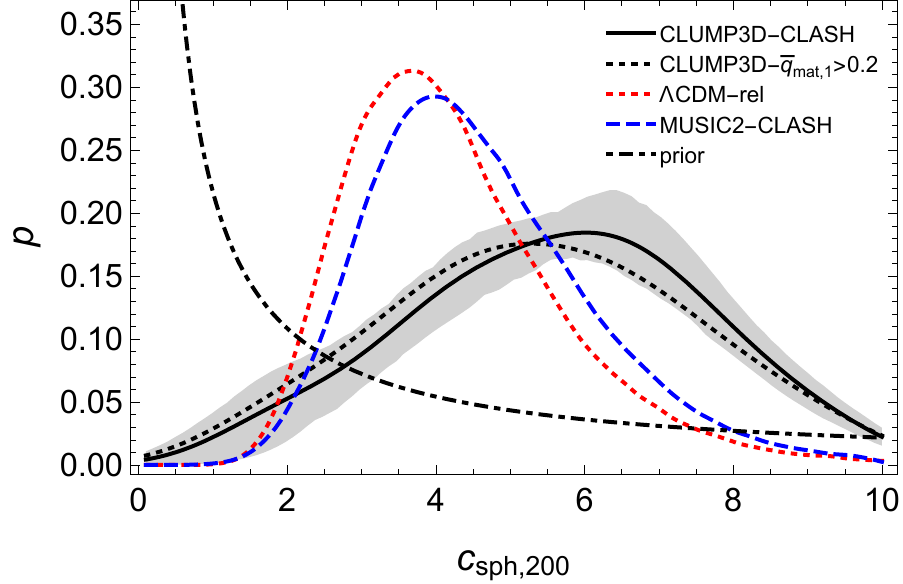}} \\
\end{tabular}
$
\end{center}
\caption{
Probability density functions: minor-to-major axis ratio (top-left panel), intermediate-to-major axis ratio (bottom-left), inclination angle $\cos \vartheta$ (top-right), and concentration (bottom-right) of the total matter distribution of the CLASH clusters. The full-black lines show the measured distribution. The shadowed regions include the 1-$\sigma$ region as obtained from a bootstrap resampling of the marginalized distributions. The black dotted lines show the distributions for the subsample with $\bar{q}_\text{mat,1}$>0.2. The red-dotted lines show the $\Lambda$CDM-rel predictions, estimated from \citet{bon+al15} given the inferred mass distribution smoothed for the observational uncertainties for the axial ratios, or the scattered mass-concentration relation of relaxed clusters from \citet{men+al14} for the concentration. The blue-dashed line is the expected distribution for the MUSIC2-CLASH simulated halos. The dot-dashed black lines show the priors. For visualisation purposes, the measured and the MUSIC2-CLASH distribution are smoothed by the Silverman's scale of the measured one-point estimates \citep{vio+al94}.
}
\label{fig_CLASH_PDF_1D}
\end{figure*}

\begin{table}
\caption{Test hypothesis. Measurements are compared to the theoretical predictions, either MUSIC2-CLASH (rows 1-5) or $\Lambda$CDM-rel (rows 6-10). For $q_\text{mat,1}$, we also report results for the subsample with $\bar{q}_\text{mat,1}>0.2$. The reported $p$-values are computed with either the Kolmogorov-Smirnov or the Pearson $\chi^2$ test. We report the mean and the standard deviation of the $p$-value distribution (cols.~2 and 4), accounting for finite sample size and observational uncertainties, and the upper limit of the 2$\sigma$-confidence region (cols.~3 and 5).
}
\label{tab_KS}
\centering
\begin{tabular}{ l r@{$\,\pm\,$}l l r@{$\,\pm\,$}l l }     
\hline
Parameter		&		\multicolumn{3}{c}{Kolmogorov-Smirnov}	& \multicolumn{3}{c}{Pearson $\chi^2$} 	\\
\hline
			&		\multicolumn{6}{c}{MUSIC2-CLASH}  \\
$c_{\text{sph},200}$	&	0.63	&	0.29	&	$\la 1$	&	0.52	&	0.25	&	0.97	\\
$q_\text{mat,1}$    	&	0.01	&	0.03	&	0.09	&	0.21	&	0.20	&	0.73	\\
$q_\text{mat,2}$    	&	0.59	&	0.31	&	$\la 1$	&	0.50	&	0.25	&	0.97	\\
$\cos\vartheta$     	&	0.40	&	0.31	&	0.95	&	0.39	&	0.24	&	0.93	\\
\multicolumn{7}{l}{Subsample $\bar{q}_\text{mat,1}>0.2$}  \\
$q_\text{mat,1}$    	&	0.07	&	0.13	&	0.59	&	0.32 &	0.23	&	0.87	\\
	&		\multicolumn{6}{c}{$\Lambda$CDM-rel}  \\
$c_{\text{sph},200}$	&	0.45	&	0.33	&	$\la 1$	&	0.41	&	0.25	&	0.94	\\
$q_\text{mat,1}$    	&	0.08	&	0.13	&	0.43	&	0.24	&	0.19	&	0.73	\\
$q_\text{mat,2}$    	&	0.53	&	0.31	&	$\la 1$	&	0.45	&	0.25	&	0.96	\\
$\cos\vartheta$     	&	0.59	&	0.31	&	$\la 1$	&	0.47	&	0.25	&	0.97	\\
\multicolumn{7}{l}{Subsample $\bar{q}_\text{mat,1}>0.2$}  \\
$q_\text{mat,1}$    	&	0.25	&	0.27	&	0.90	&	0.38	&	0.24	&	0.87	\\
\hline	
\end{tabular}
\end{table}

Cluster masses and concentrations, and shape and orientation parameters are listed in Table~\ref{tab_results}. We fitted all the CLUMP-3D parameters \citep[table~1]{ser+al17_CLUMP_M1206}, including the gas parameters, which are not reported here and will be discussed separately. 

Halo parameters are usually measured assuming spherical symmetry and by fitting the projected maps. Masses and concentrations so determined are called 2D. Here, we measure shape and concentration in 3D and we average the ellipsoidal mass profile in spherical regions. Our spherical masses are then unbiased.


The sample distributions are shown in Fig.~\ref{fig_CLASH_PDF_1D}. The theoretical $\Lambda$CDM prediction for each cluster is computed based on the observed mass distribution. For the MUSIC2-CLASH simulated clusters, we consider random samples of 16 associations, one per cluster.  The expected distribution for each cluster is convolved with a Gaussian whose dispersion is equal to the statistical uncertainty on the one-point estimate. 

The sample distributions are finally obtained by averaging the distributions of the single clusters. Measurements agree with $\Lambda$CDM predictions. For a quantitative assessment, we performed the Kolmogorov-Smirnov and the Pearson $\chi^2$ tests, see Table~\ref{tab_KS}.

Triaxial analyses facilitate the agreement of concentrations measured in massive lensing clusters with theoretical predictions \citep{ogu+al05,se+zi12}. In fact, the 3D analysis does not suffer from the orientation bias that can affect clusters preferentially elongated along the line-of-sight, whose concentration is over-estimated under the spherical hypothesis.

The CLASH clusters show a triaxial shape in good agreement with theoretical predictions. The distribution of $q_{\text{mat},1}$ slightly exceeds expectations at small values but the excess is not significant. The excess is mostly driven by three possibly unrelaxed clusters with $q_{\text{mat},1}\ls 0.2$, where the 3D combined analysis might experience some problems. M0744 and R1347 show evidence for shocks in high resolution {\it MUSTANG} SZe data \citep{kor+al11}. R1347 hosts a radio mini-halo. 
Diffuse radio emission was suspected in A2261 \citep{gia+al17}.

Adiabatic contraction and radiative cooling can make the total mass distribution rounder but AGN (Active Galactic Nuclei) feedback can mitigate the effect of cooling and make the final shape more similar to results in DM-only simulations \citep{sut+al17}. If spherically averaged profiles are considered, the baryon physics is important only within less than 10\% of the virial radius. On the other hand, the non-sphericity of the DM distribution can be affected by the baryon physics operating in the central region up to half the virial radius \citep{sut+al17}.

The MUSIC2-CLASH shapes are measured in non-radiative simulations and may differ from simulations accounting for feedback processes. In fact, the agreement is better for DM-only simulations.





X-ray regular clusters may suffer from orientation bias: prolate clusters whose major axis is aligned with the line-of-sight show round X-ray isophotes and can be preferentially included. The observed distribution of inclination angles is consistent with random orientations. The secondary peak of the distribution at high values is due to a couple of very elongated clusters as A209 which is a known merger system along the line-of-sight \citep{cas+al10}. Our algorithm can efficiently recover the orientation even in this peculiar system.

For the CLASH sample, we find $\cos \vartheta = 0.42\pm0.31$ or $\vartheta = 65\pm23\deg$. The mean angle for random orientations (MUSIC2-CLASH) is $60 \deg$ or $\cos \vartheta=0.5$ ($54 \deg$ or $\cos \vartheta=0.57$), consistent with our results and giving no evidence for any orientation bias.



\section{Conclusions}
\label{sec_concl}

We compared shapes and concentrations of X-ray selected CLASH clusters to the $\Lambda$CDM paradigm of structure formation. We performed a full three-dimensional analysis of the cluster mass profiles exploiting lensing, X-ray, and SZe measurements. We could then measure unbiased concentrations and masses and recover the halo shapes and orientations. We found that shapes and concentrations of the CLASH clusters are consistent with theoretical predictions giving a further validation of the $\Lambda$CDM paradigm.

Even though results from simulations are consistent with our measurements, we still lack a comprehensive analytical model of halo formation. The conventional ellipsoidal collapse mode and the simulations differ after the turn-around epoch \citep{sut+al16}. 
While simulated massive halos are more spherical initially, they gradually become less spherical after the turn-around epoch. This tendency is opposite to the analytical prediction \citep{sut+al16}.

Furthermore, the role of gas physics in halo shapes is yet to be fully understood  \citep{sut+al17}. In DM-only simulations, the inner regions are less spherical than the outer ones \citep{all+al06,sut+al16,veg+al17,des+al17}. Internal parts retain memory of the violent formation process keeping the major axis oriented towards the preferential direction of the infalling material while the outer regions become rounder due to continuous isotropic merging events \citep{des+al17}.

However, baryonic physics can significantly affect the non-sphericity of the DM distribution up to the half of the virial radius \citep{sut+al17}. Radiative cooling makes DM halos more spherical \citep{kaz+al04} but AGN feedback can counterbalance. As a result, the radial trend of asphericity can be opposite to that predicted  in DM-only simulations \citep{sut+al17}. 

Our results suggest that baryonic physics is not effective in making cluster rounder.

\smallskip

The authors thank Rossella Cassano and Gianfranco Brunetti for useful discussions, and Gustavo Yepes for supplying the MUSIC-2 simulations performed at the Barcelona Supercomputing Center, Spain. MS and SE acknowledge support from the contracts ASI-INAF I/009/10/0, NARO15 ASI-INAF I/037/12/0, ASI 2015-046-R.0 and ASI-INAF n.2017-14-H.0. KU acknowledges support from the Ministry of Science and Technology of Taiwan (grants 103-2112-M-001-030-MY3 and 106-2628-M-001-003-MY3) and from the Academia Sinica Investigator Award. JS was supported by NSF/AST-1617022. MM acknowledges support from MAECI and contracts ASI-INAF/I/023/12/0 and ASI n.~I/023/12/0. JV-F was supported by AYA2015-64508-P (MINECO/FEDER).






\end{document}